\documentclass[manuscript,screen]{acmart}
\AtBeginDocument{%
  \providecommand\BibTeX{{%
    \normalfont B\kern-0.5em{\scshape i\kern-0.25em b}\kern-0.8em\TeX}}}

\setcopyright{acmlicensed}
\copyrightyear{2024}
\acmYear{2024}
\acmDOI{XXXXXXX.XXXXXXX}

{\acmConference[CCAI 2024]{CHI 2024 Workshop on Child-centred AI Design}{May 11, 2024}{Honolulu, HI, USA}}
\acmBooktitle{CHI 2024 Workshop on Child-centred AI Design, May 11, 2024, Honolulu, HI, USA} 
\acmISBN{978-1-4503-XXXX-X/18/06}

\begin{document}

\title{Designing Safe and Engaging AI Experiences for Children: Towards the Definition of Best Practices in UI/UX Design}


\author{Grazia Ragone}
\email{grazia.ragone@uniba.it}
\orcid{0000-0002-8774-1789}
\affiliation{%
  \institution{University of Bari 'Aldo Moro'}
  \streetaddress{via Orabona, 4}
  \city{Bari}
  \country{Italy}
  \postcode{70125}
}

\author{Paolo Buono}
\email{paolo.buono@uniba.it}
\orcid{0000-0002-1421-3686}
\affiliation{%
  \institution{University of Bari Aldo Moro}
  \streetaddress{via Orabona, 4}
  \city{Bari}
  \country{Italy}
  \postcode{70125}
}

\author{Rosa Lanzilotti}
\email{rosa.lanzilotti@uniba.it}
\orcid{0000-0002-2039-8162}
\affiliation{%
  \institution{University of Bari Aldo Moro}
  \streetaddress{via Orabona, 4}
  \city{Bari}
  \country{Italy}
  \postcode{70125}
}
\renewcommand{\shortauthors}{Grazia Ragone} 

\begin{abstract}
This workshop proposal focuses on best practices in UI/UX design for AI applications aimed at children, emphasising safety, engagement, and ethics. It aims to address the challenge of measuring the safety, trustworthiness, and reliability of interactions between children and AI systems. Through collaborative discussions, participants will explore effective design strategies and ethical guidelines while developing methodologies for assessing the safety and reliability of AI interactions with children. This proposal seeks to foster responsible and child-centered AI design practices within the CHI community.
\end{abstract}




\received{20 February 2024}
\received[revised]{12 March 2024}
\received[accepted]{30 March 2024}

\maketitle

%\section{Author}
%\textbf{Grazia Ragone} recently awarded a doctorate in Human-Computer Interaction (HCI) from the University of Sussex (UK), where she specialised in exploring interactions with autistic children, delving into nuances such as imitation and social motor synchrony. Now positioned as a postdoc researcher at the University of Bari Aldo Moro (Italy), her focus is pioneering metrics for evaluating the symbiotic relationship between humans and AI. Her research centers on incorporating human factors to shape technology for an enhanced overall human experience.

\section{Introduction}
Creating AI systems for children necessitates a collaborative approach, drawing expertise from various fields, including psychology, education, and assessment. This proposal aims to develop a robust framework that engages diverse stakeholders in designing and testing AI systems. It ensures the systems are not only engaging and safe for children, while also flexible enough to keep up with changes in how children interact with systems.

Specifically, this proposal addresses challenges associated with measuring safety, reliability, and trustworthiness in the interaction between children and AI systems. It also delves into exploring best practices for ensuring safety, fairness, and transparency in AI user interfaces (UI) and user experiences (UX) tailored for children. By fostering discussions within the CHI community, it aims to cultivate methodologies for assessing the safety and reliability of AI interactions with children, thus advocating for responsible and child-centered AI design practices.

Central to this endeavor is a focus on fostering positive learning outcomes and engagement. The proposed framework provides guidelines and strategies to prioritise children's well-being and development in AI interface design. Through extensive stakeholder involvement and strict adherence to ethical guidelines, the goal is to facilitate the creation of AI interfaces that effectively meet the distinctive needs of young users while upholding the highest standards of safety and reliability.

To underscore the significance of creating safe, trustworthy, and reliable intelligent systems, consider an AI-driven educational game for children. Recommendations include aligning game content with educational objectives, incorporating adaptive learning features, providing real-time feedback, and ensuring accessibility. Evaluation metrics should encompass learning outcomes, knowledge retention, engagement levels, and measures of reliability, trustworthiness, and safety to safeguard the well-being of young users. 

To ensure the effectiveness and inclusivity of the proposed framework, we propose integrating innovative methodologies into the design and evaluation processes. In our suggested framework for designing AI interfaces tailored for children, we aim to create interfaces that align with children's learning needs and developmental stages. We propose integrating the following components into the framework.

Collaborative workshops involving children, educators, designers, and researchers can be organised to gather diverse perspectives and co-create design ideas \cite{Sanders08}. These workshops will serve as foundational platforms for requirement identification and brainstorming. Additionally, we suggest conducting Child-Centric Design Sessions where children actively participate in the interface design process. Their input will be solicited to ensure that the final product resonates with their preferences and needs \cite{Druin99}. Moreover, employing rapid prototyping techniques will allow us to create iterative versions of the interface \cite{Snyder03}. These prototypes will be tested with children to gather real-time feedback and refine the design based on user preferences.

Furthermore, we propose employing a combination of qualitative and quantitative research methods to gain comprehensive insights into children's interaction with the AI interface \cite{Creswell22}. This approach will involve both observational and ethnographic studies, as well as data collection, to capture nuanced user behaviors and preferences. By integrating observational methods and ethnographic research, we aim to better understand children's socio-cultural context and everyday interactions with technology \cite{Hammersley19}. By integrating these innovative methodologies, we strive to develop a robust framework for designing AI interfaces that are effective, engaging, and safe for children.

\section{Towards Ethical and Engaging AI Interfaces for Children: A Comprehensive Framework}
Designing AI interfaces for children requires a holistic approach that integrates key methodologies and practices to ensure human-centered design \cite{Schoenherr23}. Collaborative workshops involving child psychologists, educators, assessment specialists, and children themselves serve as foundational platforms for gathering insights and feedback to inform interface development. Adherence to ethical guidelines and obtaining informed consent are essential to prioritise child participants' well-being and safety \cite{Mahomed23}.

The proposed framework advocates for a series of iterative steps. We propose leveraging the opportunity of a collaborative workshop, such as the Second Workshop of Child-Centred AI Design at CHI24, to develop guidelines for designing and assessing child-centered AI systems. This workshop could serve as a platform to facilitate co-creation, requirement identification, and brainstorming of design ideas among diverse stakeholders, including child psychologists, educators, designers, and researchers.

Following the workshop, an iterative exchange of ideas and stakeholder feedback would be crucial for refining and finalising the guidelines. Input from psychologists and educators would be particularly valuable in ensuring that the guidelines prioritise positive learning outcomes and align with the developmental needs of children. By fostering collaboration and incorporating insights from various disciplines, we can develop comprehensive and effective guidelines for designing and assessing AI systems tailored to children's unique needs and preferences.

User testing sessions with children are pivotal, allowing observation of interactions to identify usability issues and understand preferences and behaviours. Involving assessment specialists ensures alignment with educational objectives and accurate evaluation of learning outcomes. Additionally, emphasising strict adherence to ethical guidelines and implementing measures to assess the symbiotic relationship between children and AI systems ensures safe and trustworthy interactions.

By following this comprehensive framework, design teams can create engaging, ethically sound AI interfaces that promote positive learning experiences for children while upholding ethical standards.

\section{Best Practices for AI UI/UX Design for Children}

Designing UI/UX for children requires careful consideration of their cognitive abilities and safety concerns. 

Key practices include the enhancement of engagement and comprehension with \textbf{a simplified design} offering clear navigation, intuitive controls, and both, \textbf{language and age-appropriate content}. Incorporating interactive features, animations, and gamification enhances the engagement and motivation of children.
The \textbf{user's customisation and personalisation} of the interface fosters autonomy and ownership. However, in the case of children, unrestricted autonomy may not always be conducive to their safety. Therefore, a parent or caregiver may need to intervene on their behalf.
Providing \textbf{clear feedback and interactive guidance} promotes independent learning. Building \textbf{trust} through \textbf{transparency} ensures that users are fully informed about how AI functions, its limitations, and how their data are utilised. We also aim to implement robust \textbf{safety measures}, including content filtering and privacy policies, to guarantee the protection of children. Inclusion should serve as the overarching key when considering diverse needs and abilities to ensure \textbf{accessibility} with inclusive design  for all children. Finally, prioritising \textbf{fairness} and \textbf{diversity} to avoid biases in content and representation should guide all designers and stakeholders within the AI community.

Following these practices, designers can create AI interfaces that offer enriching experiences while fostering children's learning and development in a digital age.

\section{Metrics for Assessing Trustworthiness, Reliability, and Safety in Human-AI Interaction}

Schneiderman's Human-Centered AI framework emphasises the importance of trustworthiness, reliability, and safety in AI interactions, with key metrics including transparency, explainability, accuracy, consistency, robustness, fairness, and user control~\cite{Shneiderman22}. For children interacting with AI interfaces, ensuring these qualities is paramount.
\emph{Transparency} ensures children understand how AI systems work, fostering comprehension and trust. \emph{Explainability} complements transparency by ensuring algorithms are understandable to young users, promoting learning. \textit{Accuracy} prevents misinformation, vital for reliable learning experiences, such as accurate translations in language learning apps.
\emph{Consistency} builds trust over time, providing reliable performance in AI-driven activities like storytelling apps. \emph{Robustness} enables adaptability, ensuring AI systems perform reliably across various inputs and scenarios, as seen in educational games adjusting to individual progress.
\emph{Fairness} guarantees equitable treatment, crucial for personalised learning platforms offering unbiased recommendations to every child. Lastly, \textit{user control} and \textit{autonomy} empower children to personalise their interactions, fostering a sense of ownership over their learning journey.
By prioritising these metrics in AI interface design and evaluation, we can create engaging, trustworthy, and safe experiences that support children's learning and development in the digital age. This systematic approach ensures positive and responsible interactions, benefiting users of all ages.

\section{Conclusion}
Efficient involvement of stakeholders, including child psychologists, educators, assessment specialists, and children themselves, is essential for designing and assessing AI interfaces that meet the unique needs of children. By adopting collaborative design approaches and incorporating feedback from diverse stakeholders, design teams can create AI interfaces that promote positive learning outcomes and engagement. Additionally, adherence to best practices for AI UI/UX design ensures safety, fairness, and transparency, fostering inclusive and empowering experiences for children interacting with AI technologies.

\section{Acknowledgement}
This research is partially supported by the co-funding of the European Union Next Generation EU: NRRP Initiative, Mission 4, Component 2, Investment 1.3 Partnerships extended to universities, research centers, companies and research D.D. MUR n. 341 of 15.03.2022 Next
Generation EU (PE0000013 “Future Artificial Intelligence Research FAIR” CUP: H97G22000210007).

\bibliographystyle{ACM-Reference-Format}
\bibliography{refs}

\end{document}